\newif\ifSUBMIT
\newif\ifCOMMENTS
\newif\ifFIGs
\newif\ifFIGoneColumn
\let\ifSUBMIT\iftrue
\let\ifCOMMENTS\iffalse
\let\ifFIGs\iftrue
\let\ifFIGoneColumn\iftrue
\documentclass{elsarticle}
\usepackage{graphicx}
\usepackage{amsmath}
\usepackage{xcolor}
\usepackage{hyperref}
\usepackage{siunitx}
\usepackage[utf8]{inputenc}
\usepackage{array,mathtools,amssymb,booktabs}
\usepackage{acronym}

\ifSUBMIT
  \ifCOMMENTS
    \usepackage{color}    
    \usepackage[normalem]{ulem}
    \def\EDITS#1{{\color{green}#1}}
    \def\STRIKE#1{{\color{red}\sout{#1}}}
    
    \def\NSTRIKE#1{{\color{blue}\sout{#1}}}
  \else
    \def\EDITS#1{#1}
    \def\STRIKE#1{}
    
    \def\NSTRIKE#1{}
  \fi
\else
  \usepackage{color}    
  \usepackage[normalem]{ulem}
 \definecolor{mygreen}{RGB}{0,180,0}    
  \def\EDITS#1{{\color{mygreen}#1}}
  \def\STRIKE#1{{\color{red}\sout{#1}}}
  
  \def\NSTRIKE#1{{\color{blue}\sout{#1}}}
\fi

\biboptions{sort&compress}

\definecolor{mygray}{RGB}{128,128,128}

\DeclareMathOperator*{\argmin}{arg\,min}

\begin{document}
\newlength\figurewide
\ifFIGoneColumn
  \figurewide=.5\columnwidth
\else
  \figurewide=.9\columnwidth
\fi

\title{Obtaining time-dependent multi-dimensional dividing surfaces 
using Lagrangian descriptors}

\author[itp1]{Matthias Feldmaier}
\author[itp1]{Andrej Junginger}
\author[itp1]{J\"org Main}
\author[itp1]{G\"unter Wunner}
\address[itp1]{%
Institut f\"ur Theoretische Physik 1, 
Universit\"at Stuttgart, 
70550 Stuttgart,
Germany}
\author[jhu]{Rigoberto Hernandez\corref{cor1}}
\cortext[cor1]{Corresponding author}
\ead{{r.hernandez@jhu.edu}}
\address[jhu]{%
  Department of Chemistry,
  Johns Hopkins University,
  Baltimore, MD 21218, USA
}

\date{\today}
\newcommand{\EQ}{Eq.}
\newcommand{\EQS}{Eqs.}
\newcommand{\FIG}{Fig.}
\newcommand{\FIGS}{Figs.}
\newcommand{\REF}{Ref.}
\newcommand{\REFS}{Refs.}
\newcommand{\SEC}{Sec.}
\newcommand{\SECS}{Secs.}
\newcommand{\eg}{e.\,g.}
\newcommand{\cf}{cf.}
\newcommand{\ie}{i.\,e.}
\newcommand{\ud}{\mathrm{d}}
\newcommand{\ue}{\mathrm{e}}
\newcommand{\kB}{k_\mathrm{B}}
\newcommand{\VLiCN}{V_\mathrm{LiCN}}
\newcommand{\VCN}{V_\mathrm{C-N}}
\newcommand{\VLi}{V_\mathrm{Li-CN}}
\renewcommand{\vec}[1]{\boldsymbol{#1}}
\newcommand{\qq}{\vec{q}}
\newcommand{\xx}{\vec{x}}
\newcommand{\vv}{\vec{v}}
\newcommand{\transpose}{\mathsf{T}}
\newcommand{\reactantpop}{\mathcal{P}}
\newcommand{\kf}{k_\mathrm{f}}
\newcommand{\etal}{\emph{et al.}}
\newcommand{\LD}{\mathcal{L}}
\newcommand{\LDf}{\LD^\text{(f)}}
\newcommand{\LDb}{\LD^\text{(b)}}
\newcommand{\LDfb}{\LD^\text{(fb)}}
\newcommand{\LDfbw}{\LD^\text{(fbw)}}
\newcommand{\Ws}{\mathcal{W}_\text{s}}
\newcommand{\Wu}{\mathcal{W}_\text{u}}
\newcommand{\Wsu}{\mathcal{W}_\text{s,u}}
\newcommand{\TSt}{\mathcal{T}}
\newcommand{\weightingf}{\chi^\text{(f)}}
\newcommand{\weightingb}{\chi^\text{(b)}}
\newcommand{\weightingfb}{\chi^\text{(f,b)}}
\newcommand{\vtherm}{v_\text{therm}}
\newcommand{\comment}[1]{\textsf{\textcolor{orange}{[#1]}}}
\newcommand{\sno}[1]{_\mathrm{#1}}
\newcommand{\no}[1]{\mathrm{#1}}
\newcommand{\acnew}[1]{\acfi{#1}\acused{#1}}
\newcommand{\atomicmass}{m_\text{u}}

\begin{abstract}\label{sec:abstract}%
Dynamics between reactants and products are often mediated by
a rate-deter\-mining barrier
and an associated dividing surface
leading to the transition state theory rate.
This
framework is challenged when the barrier is time-dependent
because its motion can give rise to recrossings across the
fixed dividing surface.
A non-recrossing
time-dependent dividing surface can neverthless be attached to the TS trajectory
resulting in recrossing-free dynamics.
We extend the formalism
---contstructed using Lagrangian Descriptors---
to systems with additional bath degrees of freedom.
The propagation of
reactant ensembles provides a numerical demonstration that our
dividing surface is recrossing-free and leads to exact TST
rates.
\end{abstract}
\begin{keyword}
Transition state theory,
Chemical reactions,
Lagrangian descriptors
\end{keyword}
\maketitle

\acrodef{DS}{dividing surface}
\acrodef{TST}{transition state theory}
\acrodef{TS}{transition state}
\acrodef{LD}{Lagrangian descriptor}
\acrodef{LDDS}{Lagrangian descriptor dividing surface}
\acrodef{PODS}{periodic orbit dividing surface}


\section{Introduction}

The accuracy in the determination of reaction rates relies on the
precision with which 
reactants and products can be distinguished 
in the underlying state space.
Usually, the boundary between
these regions contains an energetic saddle point in phase space 
to which an appropriate dividing surface (DS) can be attached.
Transition state theory (TST)
\cite{pitzer,pechukas1981,truh79,truh85,hynes85b,berne88,nitzan88,rmp90,truhlar91,truh96,truh2000,%
Komatsuzaki2001,pollak05a,Waalkens2008,hern08d,Komatsuzaki2010,hern10a,Henkelman2016}
then provides a powerful basis for the qualitative and quantitative description 
of the reaction. 
The rate is obtained from the flux through the DS and it is exact if and 
only if the DS is free of recrossings. 
Advances in the determination of this fundamental quantity can 
impact a broad range of problems in
atomic physics \cite{Jaffe00},
solid state physics \cite{Jacucci1984},
cluster formation \cite{Komatsuzaki99, Komatsuzaki02},
diffusion dynamics \cite{toller, voter02b},
cosmology \cite{Oliveira02}, 
celestial mechanics \cite{Jaffe02, Waalkens2005b},
and Bose-Einstein condensates \cite{Huepe1999, Huepe2003, Junginger2012a, 
Junginger2012b, Junginger2013b},
to name a few.

In autonomous systems, the recrossing-free DS 
is \EDITS{attached to} a normally hyperbolic invariant manifold  
that can be constructed using \eg~normal form 
expansions \cite{pollak78,pech79a,hern93b,hern94,Uzer02,Jaffe02,Teramoto11,Li06prl,Waalkens04b,Waalkens13}.
The situation becomes fundamentally different if the system is time-dependent, 
\eg~if it is driven by an external field or subject to thermal noise.
In one-dimensional time-dependent systems, a DS with the 
desired property is given by the transition state (TS) trajectory
\cite{dawn05a,dawn05b,hern06d,Kawai2009a,hern14b,hern14f,hern15a,hern16a, 
hern16h, hern16i}
which is a unique trajectory bound to the 
\EDITS{vicinity of the}
saddle for all time.

In systems with dimension greater than one,
the reacting 
particle can simply bypass the TS trajectory (point) 
by having a non-zero velocity perpendicular to 
the reaction coordinate.
Thus one must attach a multi-dimensional surface to
the TS trajectory that separates reactants and products.
The use of perturbation theory in multi-dimensional cases
provides both the TS trajectory and the associated geometry 
on which this dividing surface can be constructed.
The challenge, addressed in this Letter,
is how to obtain 
this multi-dimensional structure without perturbation theory.
One possible approach lies in the use of
the Lagrangian descriptor (LD) \cite{Mancho2010,Mancho2013}
used recently by Hernandez and Craven \cite{hern15e,hern16d}
to obtain the TS trajectory 
without resolving the DS at higher dimension.
This alternate framework is necessary when there is no useful reference 
such as in 
barrierless reactions \cite{hern16a},
and more generally to avoid the convergence
issues that invariably plague a perturbation expansion far from the reference.
In the case of field-induced ketene isomerization~\cite{hern16d},
the LD was computed across the entire phase space.
It not only revealed the structure of the DS, but also coincided with the
final state basins for each initial condition in phase space for
both 1-dimensional and 2-dimensional representations.
However, while the approach is formally applicable 
to arbitrary dimension, we have found that it is difficult
to perform the minimization of the 
naive LD, even in dimensions as low as two.

The time-dependent 
Lagrangian descriptor dividing surface (LDDS),
introduced in this Letter
is the natural extension to 
\EDITS{$n$ dimensions for $n>1$.}
\EDITS{We freely choose $2n-2$ phase-space coordinates
for which we fix the initial condidions, and use the LD approach
to identify a corresponding trajectory, which we call an anchor
trajectory.}
It is defined by the intersection of the stable and unstable manifolds
of the time-dependent 
Hamiltonian \cite{Lichtenberg82, Ott2002a}.
\EDITS{The TS trajectory is the anchor trajectory---which 
necessarily remains in the vicinity of the TS region for all past and
future time---with the least vibrational motion orthogonal to the reactive
degree of freedom.}
\EDITS{The LDDS is attached to 
the family of anchor trajectories and is
necessarly $(2n-1)$-dimensional.}
In the special case of a one-dimensional system ($n=1$),
the LDDS coincides with the 
\EDITS{moving DS on the} TS trajectory \cite{hern15e}.

\section{Theory and Methods}
\subsection{Two-Dimensional Model System}
We illustrate the construction of the LDDS by modeling the dynamics of a 
two-dimensional chemical reaction with stationary open
reactant and product basins.
Hamilton's equation of motion propagates the particle according to a 
non-autonomous Hamiltonian in mass-weighted coordinates with potential
\begin{align}
\begin{aligned}
V(x, y, t) =\;& 
E\sno{b}\,\exp\left(
  -a\left[x- \hat{x} \sin\left(\omega_x t \right)\right]^2\right) + 
\frac{\omega\sno{y}^2}{2}\left[y-\frac{2}{\pi}
\arctan\left(2 x \right)\right]^2.
\label{potential}
\end{aligned}
\end{align}
Here, $E\sno{b}$ is the height of a Gaussian barrier with width $a$ oscillating 
along the $x$ 
axis with frequency $\omega_x$ and amplitude $\hat x$, $\omega_y$ is the 
frequency of the harmonic potential in the $y$ direction, and the term 
$({2}/{\pi}) \arctan\left(2 x \right)$ is the minimum energy path whose form 
induces a nonlinear coupling between the two degrees of freedom.
For simplicity, all variables are presented in dimensionless units, where
the scales in energy (and $\kB T$), length, and time are set according to
half the maximum barrier height of the potential,
twice the variance of the Gaussian distribution,
and the inverse of the periodic frequency, respectively.
In these units, the dimensionless parameters in Eq.~\eqref{potential}
are set to
$E\sno{b}=2$, $a=1$, $\omega_x=\pi$, $\omega\sno{y}=2$, and
$\hat x = 0.4$. 

\subsection{Using Lagrangian Descriptors to Obtain Dividing Surfaces}
As the dividing surface between reactant and product basins is in general a high-dimensional hypersurface, the stable and unstable manifold itself become high-dimensional objects.
In the context of TST, 
the \ac{LD} at position $\vec x_0$, velocity $\vec 
v_0$, and time $t_0$, 
is defined as the integral \cite{hern15e,hern16a,hern16i},
\begin{equation}
\mathcal{L}(\vec{x}_0, \vec{v}_0, t_0) = \int_{t_0 - \tau}^{t_0 + \tau} 
||\vec{v}(t)||\,\no{d}t \,.
\label{LD}
\end{equation}
It is a measure of the arc length of the unique trajectory $\vec{x}(t)$ in 
forward and backward time over the time interval $[t_0-\tau;~t_0+\tau]$, and
the parameter $\tau$ is chosen such that it covers the relevant time 
scale of the system 
[in this \EDITS{letter}, we use $\tau=10$ corresponding to five periods of the 
oscillating barrier in \EQ~\eqref{potential}].
The importance of the LD~\eqref{LD} naturally results from the fact that the 
stable and unstable manifolds $\mathcal{W}\sno{s,u}$ which are attached to the 
barrier top in phase space, correspond to the minimum of the forward (f: $t_0\le 
t\le t_0 + \tau$) and backward (b: $t_0-\tau \le t \le t_0$) contributions to 
the LD,
\begin{subequations}
\begin{align}
\mathcal{W}\sno{s}(\vec{x}_0,\vec{v}_0, t_0) &= 
\argmin~\mathcal{L}^{(\no{f})}(\vec{x}_0, \vec{v_0}, 
t_0),
\label{stable}\\
\mathcal{W}\sno{u}(\vec{x}_0,\vec{v}_0, t_0) &= 
\argmin~\mathcal{L}^{(\no{b})}(\vec{x}_0, \vec{v_0}, 
t_0).
\label{unstable}
\end{align}%
\label{stable_unstable}%
\end{subequations}
Here, the function $\argmin$ denotes the argument of the 
local minimum of the \ac{LD} hypersurface close to the barrier top. 
\EDITS{In $n$ dimensions, we fix $2n-2$ variables freely
(which would be least associated with the reactive degree of
freedom) 
and perform the minimization in Eqs.~\eqref{stable_unstable}.
The intersection of these two manifolds
\begin{equation}
  \mathcal{T}(x_0,v_0,t) \equiv 
  \mathcal{W}\sno{s}(x_0,v_0,t) \cap
  \mathcal{W}\sno{u}(x_0,v_0,t), 
  \label{eq:DS-def}
\end{equation}
\EDITS{is the $t=0$ value of the}
anchor trajectory to which a moving DS can be attached.
The central result of this Letter
is that the 
\EDITS{the family of these anchor trajectories $\mathcal{T}(t)$ 
carries the associated family of moving dividing surfaces}
that we call the Lagrangian descriptor dividing surface (LDDS), 
and that we show below to be a recrossing-free DS.
The anchor surface $\mathcal{T}(t)$}
is a $(2n-2)$-dimensional object embedded in the $2n$-dimensional 
phase space meaning in the special case of a one-dimensional system,
this intersection is a single 
point, namely the position of the TS trajectory at given 
time \cite{hern15e}.

The algorithm used to obtain $\mathcal{T}(t)$ can be explained by means
of one of the insets in Fig.~\ref{fig:trajectory}. These insets show the LD 
of an
$x$-$v_x$-section for a certain time $t$ and fixed $y$ and $v_y$.
The LD is calculated according to \EQ~\eqref{LD} by integrating
trajectories with the respective initial conditions ($x, y, v_x, v_y, t$). 
They are obtained using
a standard (symplectic) Velocity-Verlet integrator with a sufficiently small
time-step to capture the time-dependence in the potential,
\EQ~\eqref{potential}, and to ensure convergence in the final 
positions and velocities.
The structure of the stable and unstable manifold is identified
through the local minima in the LD's $x$-$v_x$-section. 
Their intersection yields the phase space coordinates, $x$ and $v_x$,
of the point $\mathcal{T}(y, v_y, t)$ 
\EDITS{to which the LDDS is attached.}
Repeating this procedure for an equidistant grid in the $y$-$v_y$ space
(for a fixed time $t$)
results in a mesh of points of the LDDS $\mathcal{T}(t)$.
The smooth surfaces shown here are constructed through spline interpolation
of this mesh.

\section{Results}
\subsection{Trajectory Analysis}

\begin{figure}[t]
\includegraphics[width=\columnwidth]{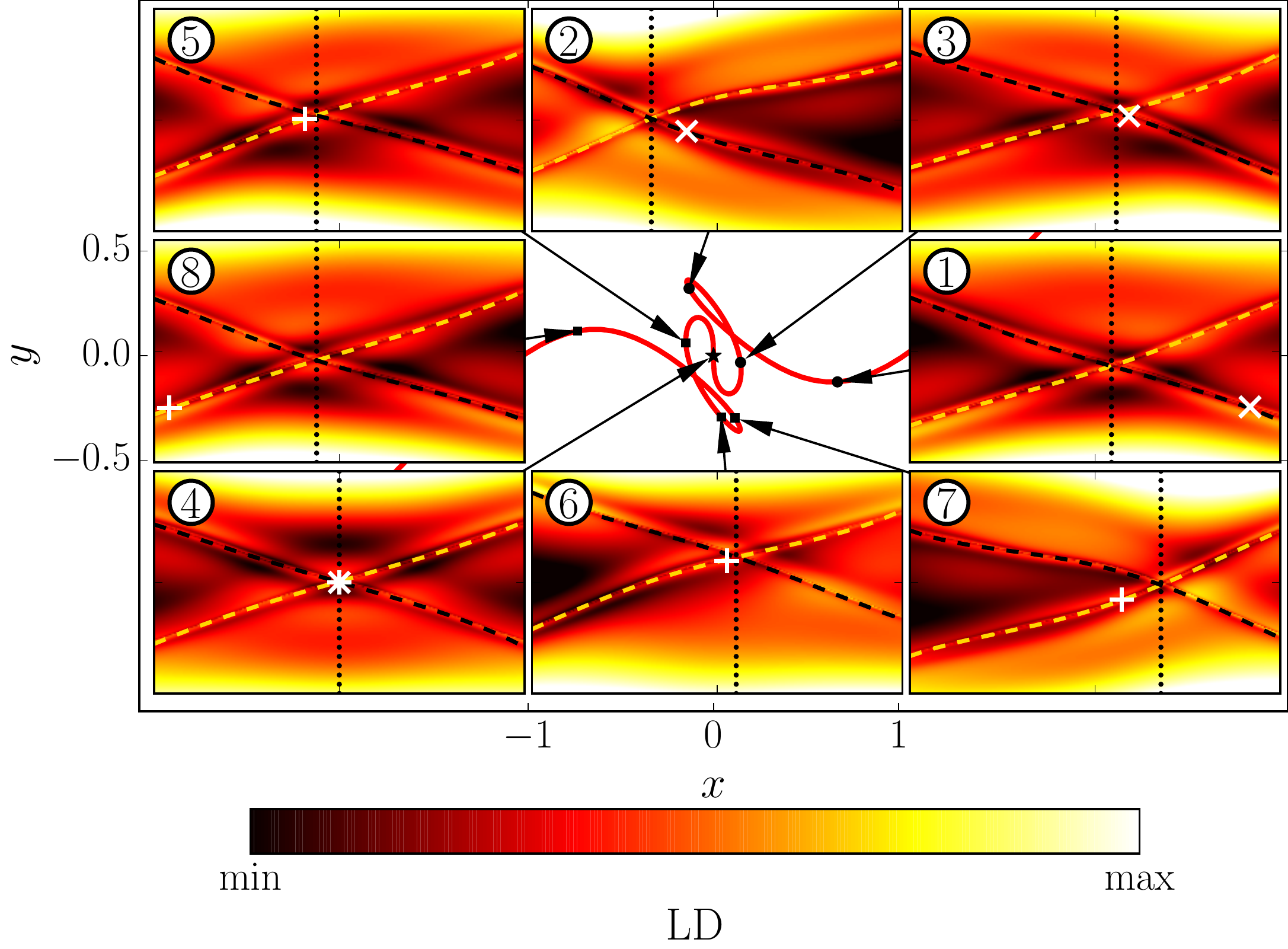}
\caption{%
An illustrative trajectory of a reactive 
particle---in red in the center panel---undergoing multiple 
oscillations in the saddle region. 
Color maps of the LD surface 
at eight selected 
points ($x, v_x$)---marked with solid circles---along
the reaction coordinate are shown at
fixed values of the bath coordinates ($y, v_y$)
displayed over the range  $-0.8 \leq x \leq 0.8$ on the horizontal 
axis and $-2.5 \leq v_x \leq 2.5$ on the vertical axis.
The stable (yellow) and unstable (black) manifolds are highlighted by dashed 
lines and the $x$-value of their intersection is marked by the vertical black 
dotted line.
The position ($x, v_x$) of the particle on a given LD surface
is shown by a white times ($\times$), asterisk ($\times\hspace{-2.5mm}+$), or 
plus ($+$) symbol, 
if the particle is located to the right, on, or to the left of
the LDDS.
}
\label{fig:trajectory}
\end{figure}

In \FIG~\ref{fig:trajectory} (center), we present a typical reactive trajectory 
(red solid line) undergoing a transition from the reactants ($x\to\infty$) to 
products ($x\to-\infty$).
Because of the oscillating barrier in the two degree of freedom 
system~\eqref{potential}, the trajectory shows several loops close to the
barrier top.
Its dominant motion is perpendicular to the reaction coordinate, but the 
trajectory also shows oscillations along the latter.
Such nontrivial oscillations are a general feature of particles with an energy 
slightly above the barrier top.
As a consequence, it is not generally
possible to define a recrossing-free DS in the configuration space alone.

Although the particle's dynamics is rather complicated near the barrier top,
the reaction dynamics becomes clearer by focusing on the 
relative motion of the particle with respect to the time-dependent manifolds.
In \FIG~\ref{fig:trajectory}, phase space portraits of the 
LD are displayed
for eight illustrative points along the selected trajectory.
The stable (unstable) manifold corresponding to 
the minimum valleys of the LD according to Eq.~\eqref{stable_unstable} 
is shown as a black (yellow) dashed line.
The time-dependent position $x^\times(y,v_y)$ where they intersect is highlighted
by a vertical, black dotted line.
In the first three points, the particle is on the RHS of $x^\times(y,v_y)$, crosses it
at point 4, and then remains on the LHS of $x^\times(y,v_y)$ for the last 4 point,
as noted with the corresponding symbol defined in the caption.
Each of the LD plots in the insets---labeled according to the corresponding
point $1,\ldots,8$---shows an $x$-$v_x$-cut through phase space for the
instantaneous values $y,v_y$ at the respective times $t$.
In this and every other trajectory we have sampled, 
the particle crosses the corresponding 
$x^\times(y,v_y)$ \EDITS{not more than} once satisfying the
recrossing-free criteria.
For a single trajectory (that fixes $y$ and $v_y$ as the two remaining
degrees of freedom \EDITS{in phase space} and therefore leads to an effective one-dimensional system),
the intersection of the manifolds~\eqref{eq:DS-def} 
thus defines a recrossing-free 
DS 
that coincides with the TS trajectory of the effective one-dimensional system.

\begin{figure}[t]
\includegraphics[width=\columnwidth]{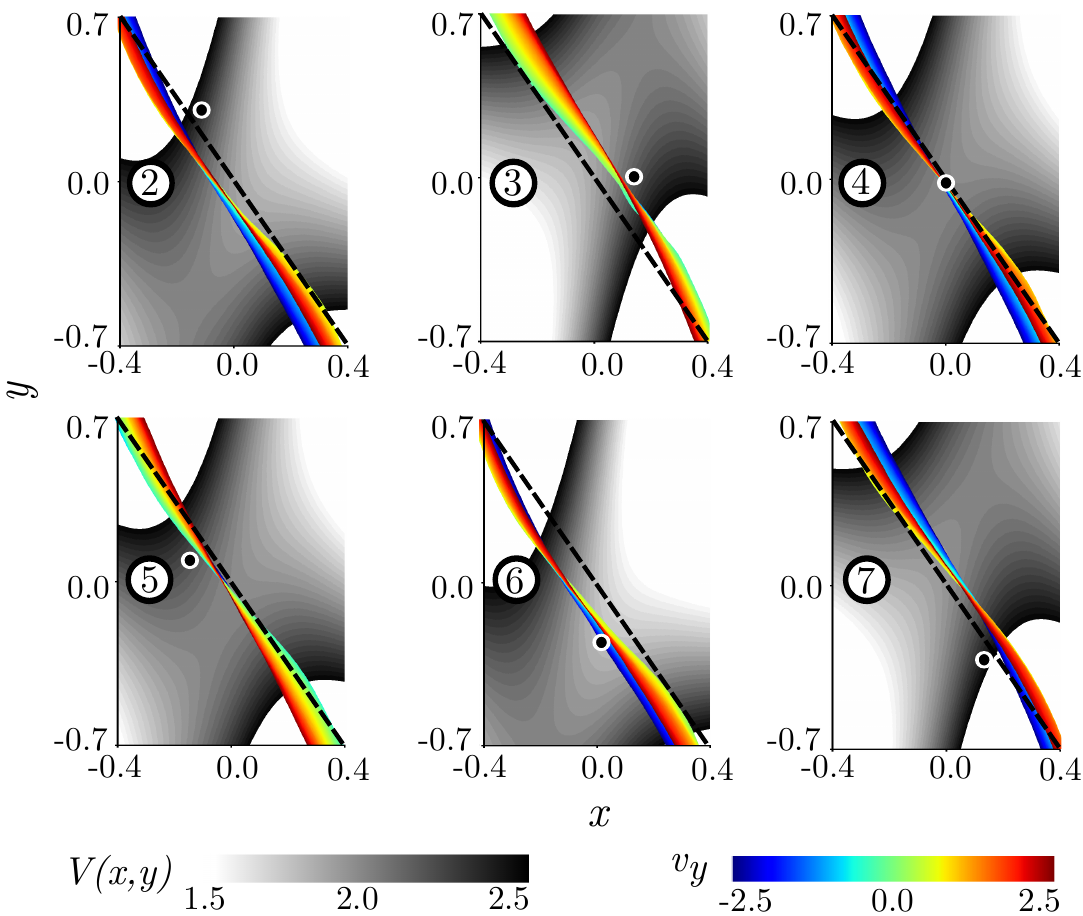}
\caption{%
Snapshots of the 
corresponding to the insets 2--7 in 
\FIG~\ref{fig:trajectory}.
The $v_y$-dimension of each 
\EDITS{$\mathcal{T}(t)$}
is encoded in the color bar on the 
right-hand side and the time-dependent saddle position is shown by the 
grayscale background.
The black dot with white circle is the instantaneous position of the reactive 
particle whose trajectory is shown in Fig.~\ref{fig:trajectory}.
The dashed black line is shown to guide the eyes and as a reference to 
emphasize the surface's motion.
}
\label{fig:several_surfaces}
\end{figure}

In the full phase space description of the two-dimensional system 
defined in Eq.~\eqref{potential}, we
can define a family of intersections $x^\times(y,v_y)$ whose
values and time-dependence vary
depending on the two remaining bath coordinates (here $y$, $v_y$)
according to Eq.~\eqref{eq:DS-def}. 
Indeed the union of these intersections
is the time-dependent, two-dimensional 
\EDITS{anchor surface to which we attach the}
LDDS. 
The 
\EDITS{$\mathcal{T}(t)$}
of Eq.~\eqref{eq:DS-def} in the $x$, $y$, $v_y$ subspace 
is displayed in Fig.~\ref{fig:several_surfaces} (panel 7) 
at the time of the corresponding point in the trajectory.
It is located near the saddle of the 
potential~\eqref{potential}, shown as a contour-surface below,
and 
exhibits a nontrivial curvature along all the axes.
Note that the calculation of the 
\EDITS{associated}
time-dependent LDDS via the intersections of the manifolds
in the $(x,v_x)$ frame for a given $y$ and $v_y$ is exemplary,
and the analogous approach using a different frame,
such as $(y,v_y)$, leads to the same result.

A higher-dimensional representation of snapshots of the 
\EDITS{anchor surface}
for the same trajectory of \FIG~\ref{fig:trajectory} is 
shown in \FIG~\ref{fig:several_surfaces}.
The nontrivial motion of the 
curved 
\EDITS{anchor}
hypersurface over time emerges as one
follows it in relation to the fixed black dashed line.
The oscillation of the 
\EDITS{anchor surface}
is in phase with that of the barrier top, 
but its amplitude is about one order of magnitude smaller 
in configuration space.
The recrossings of the
LDDS
\EDITS{attached to the anchor surface}
due to the loops in the particle's trajectory 
of \FIG~\ref{fig:trajectory} are avoided by the motion and appropriate bending
of the surface if the particle re-approaches the barrier region.

\begin{figure}[t]
\includegraphics[width=0.8\columnwidth]{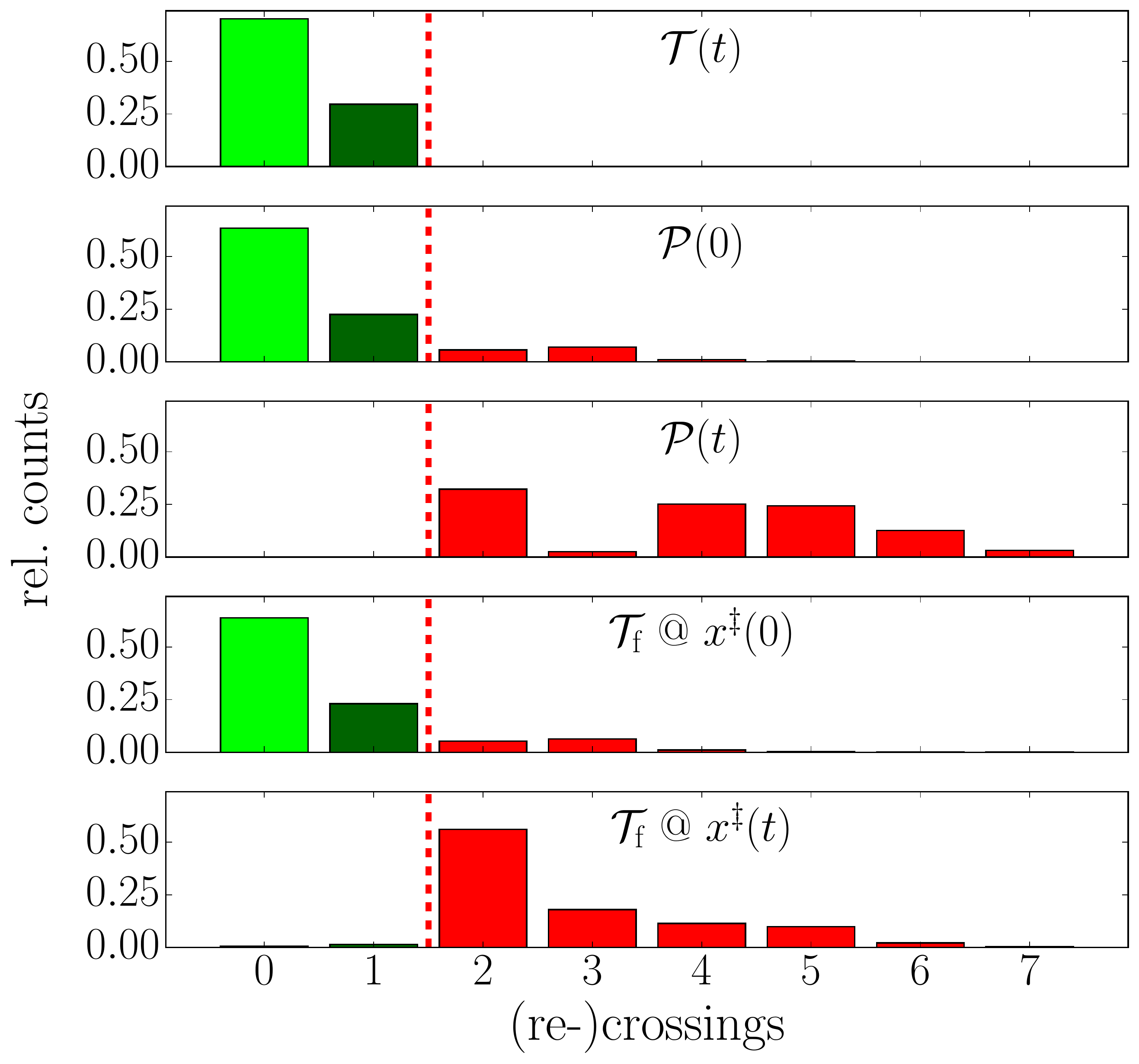}
\caption{%
Comparison of the number of (re-)crossings obtained for the same ensemble
and for different choices of DSs.
The histograms are normalized to the whole particle number.
Trajectories with 
zero and single crossings (green) 
lead to no error in the transition state estimate of the rate,
while those with two 
and more crossings (red) lead to an overestimate of the rate.
As shown in the top panel, 
\EDITS{the anchor surface}
$\mathcal{T}(t)$ 
\EDITS{to which the LDDS is attached}
does not lead to any recrossings
and gives rise to a transition state theory rate that is exact.
}
\label{fig:hists}
\end{figure}
\subsection{Ensemble Analysis}

We have, so far, demonstrated the recrossing-free nature of the LDDS 
\EDITS{attached to an anchor surface} for a 
single trajectory.
In the following, we extend this verification to an ensemble of trajectories.
Specifically, 160\,000 particles are initialized 
on an equidistant grid along the $x$-, $y$-, $v_x$-, and $v_y$-axes with 20 
points along each axis.
The grid is located close to the barrier top at $t=0$ and the grid size is 
chosen such that significant numbers of, both, reacting as well as nonreacting 
particles are observed during the time-evolution.
For this ensemble, we compare in Fig.~\ref{fig:hists} different DSs with 
respect 
to their number of (re-)crossings. As can be seen in the first panel,
the time-dependent LDDS 
\EDITS{attached to the anchor surface}
$\mathcal{T}(t)$ 
provides a recrossing-free DS 
for the whole ensemble of particles. 
For comparison, we introduce additional
DSs for which we again calculate the number 
of recrossings yielding the other panels in \FIG~\ref{fig:hists}.
A naive way to construct a DS would be a planar surface perpendicular 
to the minimum energy path and attached to the time-dependently moving
barrier top $x^\ddagger(t)$ which we refer to as $\mathcal{P}(t)$. 
In a second case, called $\mathcal{P}(0)$, we keep this planar surface fixed at
the saddle's initial position $x^\ddagger(0)$. The last two histograms are
calculated using the LDDS 
for the fixed, time-independent 
potential 
$V(x,~y,~0)$ of Eq.~\eqref{potential}. The resulting time-independent DS
is either fixed at the saddle's initial position 
[$\mathcal{T}\sno{f}~@~x^\ddagger(0)$]
or periodically moving with the top of the barrier 
[$\mathcal{T}\sno{f}~@~x^\ddagger(t)$].
As \FIG~\ref{fig:hists} shows, the LDDS is the only choice for which either no 
or only single crossings occur while two and more crossings do not occur.
By contrast, all the other choices of DSs exhibit several recrossings 
and will therefore lead to overestimates in the corresponding rates.

\begin{figure}[t]
 \includegraphics[width=\columnwidth]{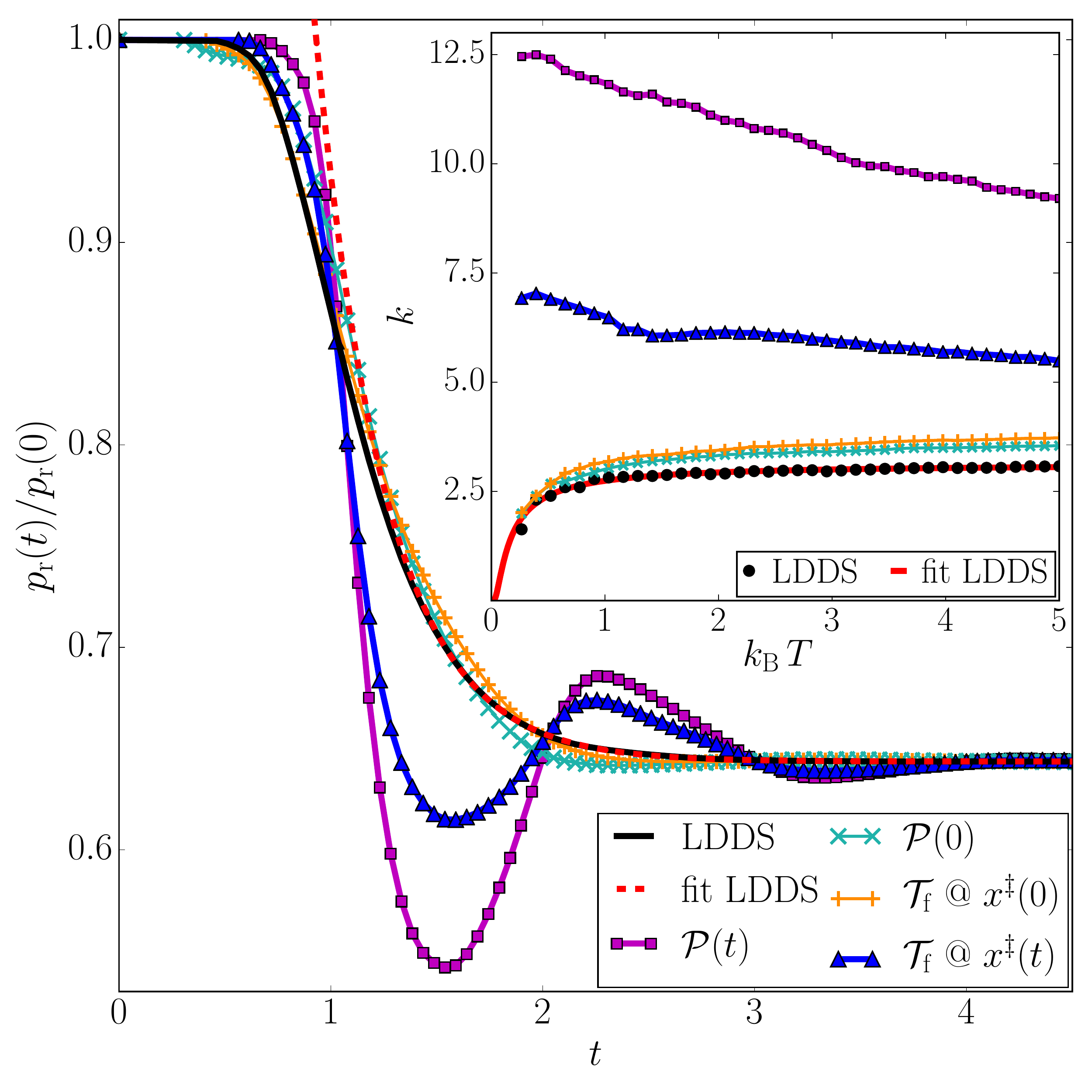}
\caption{%
Black line: relative number of reactants $p\sno{r}$ over time for an 
initial thermal ensemble ($k\sno{B}T = 5.13$) of reactant particles crossing 
the 
LDDS. 
The corresponding curve (red dashed line) yields a rate of $k=3.03$ via an 
exponential fit. 
The other curves show the number of reactants for those surfaces mentioned in 
\FIG~\ref{fig:hists}. 
The inset shows the temperature dependence of the rates $k$ obtained from the 
fit \eqref{eq:fit-exponential}.
The rates obtained from the LDDS yields a high-temperature rate of $k_\infty = 
3.14$. 
See text for further description. 
}
\label{fig:rate_calculation}
\end{figure}

Finally, we regard a thermal ensemble of $10^6$ particles in the reactant 
well with the density distribution
\begin{equation}
\rho(\vec{x}, \vec{v}) = \rho\sno{therm}\,\delta(x+2)\,\Theta(v_x) \,,
\end{equation}
where $\rho\sno{therm}$ is a Boltzmann-distribution, $\delta$ is the 
Dirac-delta 
function and $\Theta$ the Heaviside step function. 
The corresponding time-evolution of the reactant population $p\sno{r}(t)$
is shown in Fig.~\ref{fig:rate_calculation}.
The decay corresponding to the LDDS is the only strictly monotonic one.
Both of the time-independent DSs
exhibit a modulation of the 
reactant population with a very small but nonvanishing amplitude.
When fixed DSs are attached to the barrier top, $x^\ddagger(t)$, the 
oscillations are huge.
Each increase in the number of reactants is due to recrossings through the 
respective surface. 
Note that for $t\rightarrow \infty$, the reactant population for all DSs is the 
same, as the particles have fallen down from the barrier either in the reactant 
or in the product basin and their classification is independent of the 
choice of the dividing surface, as long as it is located sufficiently close to 
the saddle.

The red dashed line in Fig.~\ref{fig:rate_calculation} shows an exponential 
fit 
\begin{equation}
p\sno{r}(t) = p\sno{r,0}\,\no{e}^{-kt} + c
\label{eq:fit-exponential}
\end{equation}
to the long-time decay of the reactant population
with fit coefficients $p\sno{r,0}, k, c$ from which we extract the reaction rate 
$k$.
The respective rates obtained for the different DSs 
at various temperatures $k\sno{B}T$ are shown in the inset of 
Fig.~\ref{fig:rate_calculation}.
Rates obtained from the LDDS are the smallest throughout 
as should be
expected from a recrossing-free DS.
The time-independent $\mathcal{P}(0)$ and $\mathcal{T}\sno{f}(0)$ 
yield slightly higher rates while the 
rates obtained from the DSs attached to the barrier top [$\mathcal{P}(t)$ and 
$\mathcal{T}\sno{f}(t)$] overestimate the rates by a factor of 2 to 5 due to 
recrossings.
In addition, the LDDS rates exhibit a temperature dependence according to 
Arrhenius' rate
equation (see red fit curve in the inset of Fig.~\ref{fig:rate_calculation}),
\begin{equation}
k(k\sno{B}T) = 
k_\infty\,\exp\left(-\frac{\Delta E\sno{eff}}{k\sno{B}T}\right),
\end{equation}
where $\Delta E\sno{eff}=0.135$ is the effective height of the potential 
and $k_\infty = 3.14$ is the high-temperature limit of the rate.
The effective barrier height is significantly lower than the spread of 
barrier heights between 1.70 to 2.00 at the naive saddle point. 
The fact that $\Delta E \sno{eff}$
is not higher than these barrier heights is a good consistency check.
It is lower in energy because the driving of the system
maintains the effective reactant population
in an activated state energetically higher than the
naive reactant population near the minimum of the potential.

\section{Conclusion}

In this Letter,
we have developed and verified the 
explicit construction of the time-dependent LDDS 
for multidimensional systems as a recrossing-free dividing surface.
It reduces to the well-known TS trajectory \cite{dawn05a}
formalism in the one-dimensional limit.

The central result of this Letter is thus the justification and validation
of a generalization of the 
\EDITS{LDDS}
method whose underlying
minimization converges even in higher dimensions.
Specifically, the construction is realizable 
because the LD \eqref{LD} 
remains a scalar quantity regardless of the phase space dimension.
The construction also results in a global DS, 
\ie~the recrossing-free property does not only hold close 
to the barrier top but for the complete hypersurface in full phase space.
The method thus positions us to resolve the dividing
surface (and associated reaction rates)
in time-dependent 
molecular reactions
\cite{Yamanouchi2002,Sussman2006,Kawai11laser,Keshavamurthy2009,
Keshavamurthy2015,Revuelta2015},
and perhaps also spintronic devices 
\cite{Taniguchi2013,Apalkov2005}
driven by tailored external fields or thermal noise.

\section*{Acknowledgments}
AJ acknowledges the Alexander von Humboldt Foundation, Germany, 
for support through a Feodor Lynen Fellowship.
RH's contribution to this work was supported by 
the National Science Foundation (NSF) through Grant 
No.~CHE-1700749.
This collaboration has also benefited from support
by the people mobility programs, and most recently by the
European Union’s Horizon 2020 research and innovation
programme under Grant Agreement No.~734557.
\EDITS{Surface} plots 
\EDITS{have been} made with the \emph{Mayavi} software 
package \cite{ramachandran2011mayavi}.

\section*{References}
\bibliography{p98paper}
\end{document}